\pdfoutput=1
\documentclass[aps,prd,amsmath,floats,floatfix, twocolumn,
superscriptaddress,nofootinbib,showpacs,longbibliography]{revtex4-1}

\usepackage[T1]{fontenc}
\usepackage[utf8]{inputenc}
\usepackage{lmodern}

\usepackage{verbatim}
\usepackage{makecell}

\usepackage[dvipsnames, usenames]{xcolor}
\definecolor{linkcolor}{rgb}{0.0,0.3,0.5}
\usepackage[hypertexnames=false, unicode, colorlinks=true, linkcolor=linkcolor,
citecolor=linkcolor, filecolor=linkcolor,urlcolor=linkcolor,
pdfusetitle]{hyperref}

\usepackage[all]{hypcap}
\usepackage{graphicx}
\usepackage{xspace}
\usepackage{braket}
\usepackage{amssymb}
\usepackage[normalem]{ulem} 
\usepackage{bm} 

\usepackage{microtype}

\usepackage{blindtext}
\usepackage{mathrsfs}
\usepackage{orcidlink}
\graphicspath{%
  {figs/}%
}

\DeclareMathAlphabet{\mathpzc}{OT1}{pzc}{m}{it}

\newcommand{\h}{\mathpzc{h}}
\newcommand{\hlm}{\mathpzc{h}_{\ell m}}
\newcommand{\omegaorb}{\omega_{\mathrm{orb}}}
\newcommand{\flow}{f_{\mathrm{low}}}
\newcommand{\chieff}{\chi_{\mathrm{eff}}}

\newcommand{\chiAz}{\chi_{1z}}
\newcommand{\chiBz}{\chi_{2z}}

\newcommand{\bfemph}[1]{\emph{\textbf{#1}}}

\newcommand{\NRSurOld}{\texttt{NRHybSur3dq8}\xspace}
\newcommand{\NRSurNew}{\texttt{NRHybSur3dq8\_CCE}\xspace}

\begin{document}

\title{Numerical relativity surrogate model with\\memory effects and post-Newtonian hybridization}

 \newcommand{\Cornell}{\affiliation{Cornell Center for Astrophysics and Planetary
     Science, Cornell University, Ithaca, New York 14853, USA}}
  \newcommand{\Caltech}{\affiliation{Theoretical Astrophysics 350-17, California
 Institute of Technology, Pasadena, CA 91125, USA}}
  \newcommand{\MaxPlanck}{\affiliation{Max Planck Institute for Gravitational
     Physics (Albert Einstein Institute), Am M{\"u}hlenberg 1, D-14476 Potsdam,
     Germany}}
\newcommand{\UMassD}{\affiliation{Department of Mathematics,
    Center for Scientific Computing and Data Science Research,
    University of Massachusetts, Dartmouth, MA 02747, USA}}
\newcommand{\UMiss}{\affiliation{Department of Physics and Astronomy, 
University of Mississippi, University, Mississippi 38677, USA}}

\author{Jooheon Yoo \orcidlink{0000-0002-3251-0924}}\Cornell
\author{Keefe Mitman \orcidlink{0000-0003-0276-3856}}\Caltech
\author{Vijay Varma \orcidlink{0000-0002-9994-1761}}\MaxPlanck
\author{Michael Boyle \orcidlink{0000-0002-5075-5116}}\Cornell
\author{Scott E. Field \orcidlink{0000-0002-6037-3277}}\UMassD
\author{\\Nils Deppe \orcidlink{0000-0003-4557-4115}} \Caltech
\author{Fran\c{c}ois H\'{e}bert \orcidlink{0000-0001-9009-6955}} \Caltech
\author{Lawrence E.~Kidder \orcidlink{0000-0001-5392-7342}} \Cornell
\author{Jordan Moxon \orcidlink{0000-0001-9891-8677}} \Caltech
\author{Harald P. Pfeiffer \orcidlink{0000-0001-9288-519X}}\MaxPlanck
\author{Mark A. Scheel \orcidlink{0000-0001-6656-9134}}\Caltech
\author{Leo C. Stein \orcidlink{0000-0001-7559-9597}}\UMiss
\author{Saul A. Teukolsky \orcidlink{0000-0001-9765-4526}}\Cornell \Caltech
\author{William Throwe \orcidlink{0000-0001-5059-4378}} \Cornell
\author{Nils L.~Vu \orcidlink{0000-0002-5767-3949}} \Caltech \MaxPlanck

\hypersetup{pdfauthor={Yoo et al.}}
\date{\today}

\begin{abstract}
Numerical relativity simulations provide the most precise
templates for the gravitational waves produced by binary black hole mergers.
However,
many of these simulations use an incomplete waveform extraction
technique---extrapolation---that fails to capture important physics, such as gravitational memory effects.
Cauchy-characteristic evolution (CCE), by contrast, is a much more physically accurate extraction
procedure that fully evolves Einstein's equations to future null infinity and
accurately captures the expected physics. In this work, we present a new surrogate model, \NRSurNew, built from CCE waveforms that have been mapped to the post-Newtonian (PN) BMS
frame and then hybridized with PN and effective one-body (EOB) waveforms. This model is trained
on 102 waveforms with mass ratios $q\leq8$ and aligned spins
$\chiAz, \, \chiBz \in \left[-0.8, 0.8\right]$. The
model spans the entire LIGO-Virgo-KAGRA (LVK) frequency band (with
$f_{\text{low}}=20\text{Hz}$) for total masses $M\gtrsim2.25M_{\odot}$ and
includes the $\ell\leq4$ and $(\ell,m)=(5,5)$ spin-weight $-2$
spherical harmonic modes, but not the $(3,1)$, $(4,2)$ or $(4,1)$ modes. We find that \NRSurNew can accurately reproduce the
training waveforms with mismatches $\lesssim2\times10^{-4}$ for total masses
$2.25M_{\odot}\leq M\leq300M_{\odot}$ and can, for a modest degree of extrapolation, capably model outside of its training region. 
Most importantly, unlike previous waveform models,
the new surrogate model successfully captures
memory effects.
\end{abstract}

\maketitle

\section{Introduction}
\label{sec:introduction}

To date, there have been a total of 90 joint detections of gravitational wave (GW) signals by the LIGO\footnote{The Laser Interferometer Gravitational-Wave Observatory.}~\cite{TheLIGOScientific:2014jea} and Virgo~\cite{TheVirgo:2014hva} collaborations. But, with increased sensitivity in future observation runs and the inclusion of KAGRA\footnote{The Kamioka Gravitational Wave Detector.}~\cite{KAGRA:2020tym} as well as other
proposed future detectors, such as the Einstein Telescope~\cite{Punturo:2010zz}, the Cosmic Explorer~\cite{Reitze:2019iox}, and the space-based LISA\footnote{The Laser Interferometer Space Antenna.}~\cite{LISA:2017pwj}, 
the number of gravitational wave observations is expected 
to increase dramatically~\cite{Aasi:2013wya,Evans:2016mbw}.
To fully take advantage of the ever-expanding catalog of gravitational wave signals from compact binaries,
it is crucial that we have high-fidelity waveform templates
to compare the observed signals to. This is because accurate waveform templates are necessary for reliably 
extracting astrophysical source properties that provide important information about the binaries' formation channels and also for performing unique tests of general relativity.

\begin{figure*}[!t]
	\centering
	\includegraphics[width=1.0\textwidth]{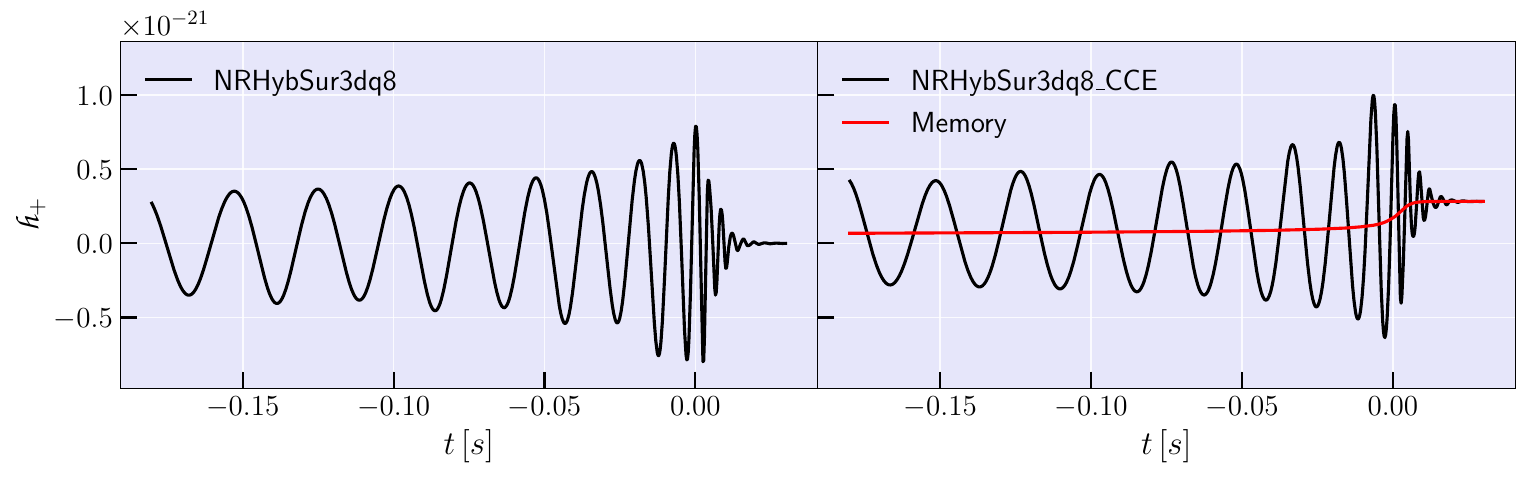}
        \caption{Plus polarization of the strain for a GW150914-like event computed using the previous surrogate \NRSurOld (left) and the new surrogate \NRSurNew (right). The exact parameters that are used to calculate these two waveforms are $m_{1}=36M_{\odot}$, $m_{2}=29M_{\odot}$, $\chiAz=0.32$, $\chiBz=-0.44$, $D_{L}=410\mathrm{Mpc}$, $\iota=\pi/2$, and $\varphi=0$. Because we wish to highlight the main difference between these surrogates, we used an inclination angle of $\iota=\pi/2$, i.e., the ``edge on'' orientation, for which the memory (red) (computed using Eq. (17b) of Ref.~\cite{Mitman:2020bjf}) is maximized. Note that the memory roughly scales as $\sin^2(\iota)$.}
	\label{fig:GW150914}
\end{figure*} 

Numerical relativity (NR) is the only 
\textit{ab initio} method for solving Einstein's 
equations for the coalescence of two compact objects and has played
a fundamental role in both GW theory and GW astronomy~\cite{Pretorius:2005gq,
Campanelli:2005dd,Baker:2005vv,SXSCatalog}. Even so,
despite continued efforts by the NR community to make simulations more computationally efficient, they
are still prohibitively expensive for key multi-query applications, such as
parameter estimation. Because of this bottleneck, numerous waveform models have
been developed~\cite{Ossokine:2020kjp,Khan:2019kot,Pratten:2020ceb,Estelles:2021gvs,Nagar:2020pcj,Akcay:2020qrj} that can be evaluated
much faster than evolving an entire NR simulation. By
construction, these semi-analytical models rely
on physically-motivated or phenomenological 
assumptions to reduce the complexity of parameter space. They then calibrate the remaining free parameters by comparing to the waveforms produced by NR simulations. While these waveform models tend to be fast enough for performing GW data analyses, they are not nearly as accurate or reliable as NR waveforms.

NR surrogate models are a more recent addition to the collection of compact
binary waveform
models~\cite{Blackman:2015pia,Blackman:2017dfb,Blackman:2017pcm,Varma:2018mmi,Varma:2019csw,Islam:2021mha,Yoo:2022erv}.
Unlike semi-analytic models, NR surrogates instead take a data-driven approach:
training the waveform model directly on the waveforms output by simulations
without the need to make any assumptions about the physics. Consequently, these
surrogates recover NR waveforms much more faithfully than other semi-analytical
models. However, because of this unique data-driven approach to waveform modeling, NR surrogate models
can only be constructed for the regions of parameter space in which NR simulations
exist.\footnote{Surrogate models have also been constructed for semi-analytical EOB models~\cite{Purrer:2014fza,Gadre:2022sed}.}

Both NR surrogates and other semi-analytical models have played a crucial role in studying 
previous detections of gravitational waves~\cite{LIGOScientific:2020tif,LIGOScientific:2021usb,LIGOScientific:2021djp,LIGOScientific:2020iuh,Varma:2021csh,Varma:2021xbh,Varma:2022pld,Varma:2020nbm}. 
Up until now, however, every waveform model has always either been trained or calibrated by working with `extrapolated' waveforms.
Because NR simulations of binaries are necessarily run in a finite volume, one needs a method of `extracting' the asymptotic waveform 
from the finite-volume data that is produced by the simulation. This is because the gravitational radiation that we observe on Earth can be well-approximated by the radiative solutions to Einstein's equations at future null infinity. In NR, an `extrapolated' waveform 
refers to the asymptotic waveform from an extraction procedure called extrapolation.\footnote{Extrapolation uses Regge-Wheeler-Zerilli extraction to compute the strain waveform on a series of concentric spheres of constant coordinate radius and then extrapolates these values to future null infinity by fitting a power series in $1/r$~\cite{Boyle:2009vi,Boyle:2019kee,Bishop:2016lgv}.}
However, a major limitation of these extrapolated waveforms, which are the waveforms
currently included in the SXS~\cite{SXSCatalog,Boyle:2019kee}
and other catalogs~\cite{Jani:2016wkt, Healy:2017psd}, is that they do
not accurately capture phenomena known as memory
effects~\cite{Zeldovich:1974gvh,1987Natur.327..123B,Thorne,Christodoulou}.

Gravitational memory effects correspond to persistent phenomena that two
observers can measure after the passage of gravitational radiation~\cite{Zeldovich:1974gvh,1987Natur.327..123B,Christodoulou,Thorne}. In particular, to measure memory effects, observers must measure the change in spacetime induced by the passage of radiation between two times: one before and one after the radiation. While there are
several types of memory effects~\cite{Compere:2019odm,Grant:2021hga}, the two most
prominent, and therefore detectable, effects are the
displacement~\cite{Zeldovich:1974gvh,1987Natur.327..123B,Thorne,Christodoulou}
and the spin~\cite{Pasterski:2015tva} memories. The displacement memory is what
two initially comoving observers will measure, while the spin memory is what
two observers with initial relative velocities will
measure, in conjunction with the usual displacement that they experience due to their nonzero relative velocities~\cite{Grant:2021hga}.

Apart from this classification of memory by the changes experienced by observers, there is also a classification in the way in which memory is sourced: ordinary and null.\footnote{Originally this classification was linear and nonlinear~\cite{Zeldovich:1974gvh,1987Natur.327..123B,Thorne}, but this terminology changed in recent years to more accurately reflect the physics sourcing this phenomenon~\cite{Bieri:2013ada}.} Ordinary memory refers to the memory that is sourced by changes in the $\ell\geq2$ mass multipole moment of ``ordinary'' unbound masses, while null memory refers to the memory that is sourced by a change in the energy radiated per unit solid angle due to the radiation of ``null'' gravitons. Consequently, null memory can be thought of as a form of the ordinary memory in which the unbound masses are individual gravitons. Generally, ordinary memory will be most prominent in unbound scattering processes, such as hyperbolic black hole encounters~\cite{1977ApJ.216.610T,Damour:2014afa,Cho:2018upo,Nagar:2020xsk,Damour:2022ybd,Rettegno:2023ghr}. In contrast, null memory will be most prominent in bound scattering processes, such as binary black hole mergers. Because of this, in this work we will primarily be interested in the null memory.

While memory effects are an undetected phenomenon, various works have investigated their
detectability using a forecast of future binary merger
observations~\cite{Hubner:2019sly,Boersma:2020gxx,Grant:2022bla} and their observational consequences~\cite{Ebersold:2020zah,Tiwari:2021gfl,Lopez:2023aja}. Furthermore, apart from their allure as a means to test Einstein's theory of relativity,
memory effects have also attracted significant attention in the theory community
because of their inherent connection to asymptotic symmetries and soft
theorems~\cite{Strominger:2014pwa,He:2014laa,Weinberg:1965nx,Pasterski:2015tva,Compere:2019odm,Haco:2018ske,Hawking:2016msc}.
Accordingly, it is crucial that templates for gravitational waves contain
memory.

While extrapolated NR waveforms ultimately fail to correctly capture memory
effects,\footnote{%
The reason why memory effects are not correctly resolved in extrapolated waveforms is because memory effects tend to have a much longer spatial dependence than the oscillatory components of the waveform. Consequently, when expressing the waveform as a series of $1/r$ terms, the convergence of the memory's contribution to the waveform is very slow and hard to capture~\cite{Boyle:2009vi,Boyle:2019kee,Bishop:2016lgv}.}\textsuperscript{,}\footnote{%
While some of the missing memory in extrapolated
waveforms can be computed through post-processing~\cite{Talbot:2018sgr,Mitman:2020bjf}, there are
certain types of memory effects that cannot be corrected, e.g., spin memory,
which makes extrapolated waveforms impractical for formal analyses of memory.} there is a much more
robust type of asymptotic waveform extraction, called Cauchy-characteristic
evolution (CCE), which fully evolves Einstein's equations to future null
infinity and correctly resolves the various memory
effects~\cite{Bishop:1996gt,Moxon:2020gha,Moxon:2021gbv,Mitman:2020pbt}. 

In this work, we build a hybridized NR surrogate model, \NRSurNew, which is
trained on CCE waveforms that have been created using the CCE module of the
code \texttt{SpECTRE}~\cite{spectrecode,Moxon:2020gha,Moxon:2021gbv}. Furthermore, to build
our hybrid waveforms, using the technique outlined in~\cite{Mitman:2022kwt}, we
first map our NR waveforms to the post-Newtonian (PN) BMS
frame~\cite{Mitman:2021xkq,Mitman:2022kwt} before we hybridize them with effective one-body (EOB) phase-corrected PN
waveforms. By doing so, we find that \NRSurNew performs on par with a related
extrapolated version of this surrogate, \NRSurOld, and in certain scenarios
even outperforms \NRSurOld, all while containing previously unresolved physical
effects. As an illustration of our waveform model, we provide Fig.~\ref{fig:GW150914},
which shows the correct waveform for a GW150914-like binary black hole merger event. 

Throughout this paper, we adopt the common notations used by previous works.
The mass ratio is denoted as $q=m_{1}/m_{2}$, where $m_{1}$ ($m_{2})$ denotes the 
mass of heavier (lighter) black hole, while the aligned spin of the heavier (lighter) black hole (in the direction of the binary's orbital angular momentum) is denoted as $\chi_{1z}$ ($\chi_{2z}$). We use $D_{L}$ to denote the luminosity distance, $\iota$ to denote the inclination angle between the orbital angular momentum and the line-of-sight to the detector, and $\varphi$ to denote the azimuthal angle. Furthermore, when outputting our waveform templates, we represent the two polarizations of the gravitational wave---the plus and cross polarizations---as a single complex waveform, $\h = \h_{+} - i\h_{\times}$, which we then further decompose into a sum of spin-weight $-2$ spherical harmonic modes denoted as $\hlm$: 
\begin{equation}
\label{eqn:hlm}
\h(t,\iota,\varphi) = \sum^{\infty}_{\ell=2} \sum^{\ell}_{m=-\ell} \hlm(t)\, {}_{-2}Y_{\ell m}(\iota,\varphi).
\end{equation}
Here ${}_{-2}Y_{\ell m}$ are the spin-weight $-2$ spherical harmonics.
In Eq.~\eqref{eqn:hlm}, the quadrupole modes ($\ell=\lvert m \rvert= 2$)
typically dominate the sum; however, the other modes
are also important for estimating binary source properties~\cite{Varma:2016dnf,
Varma:2014jxa, Capano:2013raa, Shaik:2019dym, Islam:2021zee}. Therefore,
our new model \NRSurNew includes $\ell\leq4$ and $(5,5)$ spin-weighted
spherical harmonic modes, but not the $(3,1)$, $(4,2)$ and $(4,1)$ modes. The reason for excluding these three modes
is explained in Appendix~\ref{sec:challenge}. Like its predecessor, the new model
\NRSurNew is an aligned-spin model, restricted to binary black holes (BBHs)
whose spins are aligned with the system's orbital angular momentum. 
Thus, due to orbital-plane symmetry,
we do not have to model the $m<0$ modes separately as they can be obtained from $m>0$ through the well-known relation $\h_{\ell\, (-m)} = (-1)^{\ell} \hlm^{*}$, where $\hlm^{*}$ represents the complex conjugate of $\hlm$.

The rest of the paper is organized as follows. In Sec.~\ref{sec:methods}, we describe 
the entire construction of \NRSurNew. In Sec.~\ref{sec:errors}, we then
evaluate the errors involved in building this model. In particular, we check the error due to hybridization, the error of the surrogate itself, the success of the surrogate in extrapolating to values outside its training range, and the difference between \NRSurOld and \NRSurNew. Finally, in Sec.~\ref{sec:conclusion}, we conclude with a few closing remarks. \NRSurNew, our new surrogate model, has been made publicly available
through the python package \texttt{gwsurrogate}~\cite{gwsurrogate}.

\section{Methods}
\label{sec:methods}

In this section, we outline the steps that are required for building the new surrogate model
\NRSurNew. More specifically, in the subsequent text, we discuss the parameter space that our training waveforms will cover, Bondi-van der Burg-Metzner-Sachs 
(BMS) frame fixing, hybridization, and, finally, the routine for constructing the surrogate model \NRSurNew.

\subsection{Training set generation}
\label{sec:training_set_generation}

\begin{figure}[t]
\centering
\includegraphics[width=0.5\textwidth]{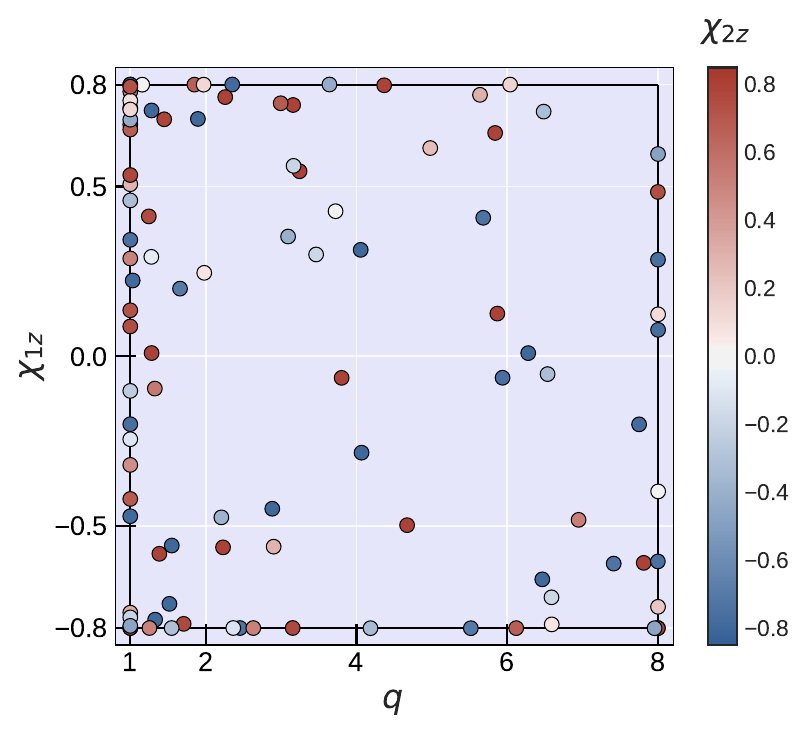}
\caption{
Training set parameters used in the construction of the new surrogate model \NRSurNew.
There are a total of 102 training data points used for \NRSurNew, which exactly match those of \NRSurOld minus 
two points, for which the initial world tube data for CCE was not available.
The boundary of the training region is represented with the black rectangle: 
$1\leq q \leq 8$ and $-0.8 \leq \chiAz ,\, \chiBz \leq 0.8$. }
\label{fig:paramspace}
\end{figure}

To build the new surrogate model \NRSurNew, we need a set of training waveforms in addition to their corresponding binary
parameters. One cannot know, a priori, the optimal distribution of binary parameters
for training the surrogate model. Fortunately, a previous surrogate model, \NRSurOld, already explored the parameter space that we are interested in: mass ratio $q \in [1,8]$ and $\lvert \chiAz \rvert,\,
\lvert \chiBz \rvert \leq 0.8$, where $\chiAz$ ($\chiBz$) is the spin of the heavier (lighter) black hole in the direction of the orbital angular momentum~\cite{Varma:2018mmi}.
Hence, we use the same set of existing NR simulations (SXS:BBH:1419--1509, but not SXS:BBH:1468 or SXS:BBH:1488)
that was used for training \NRSurOld.
For an equal mass simulation with unequal spins,
we can exchange the two BHs to obtain an extra training data point. 
This is performed by applying a rotation (along the z-axis, defined as the axis of the orbital angular momentum of the BBH) by $\pi$ 
to the waveform of $(q,\chiAz,\chiBz) = (1, \chi,\tilde{\chi})$
to obtain an extra waveform corresponding to $(q,\chiAz,\chiBz) = (1, \tilde{\chi},\chi)$ for $\chi\neq\tilde{\chi}$.
From the above 89 NR simulations, there are 13 of these cases, leading to 102 distinct training data.\footnote{Note that two NR simulations (SXS:BBH:1468 and 1488) are missing the world tube data 
that is necessary to produce the CCE waveforms of interest. This is why in this work we have 102 training data rather than the 104 training data used in~\cite{Varma:2018mmi}.} In Fig.~\ref{fig:paramspace}, we show the three-dimensional distribution ($q,\chiAz,\chiBz$) of our training parameters.

For each NR simulation used in our model, we extract
the asymptotic waveform at future null infinity $\mathscr{I}^{+}$ using the \texttt{SpECTRE} code's implementation of CCE~\cite{spectrecode,Moxon:2021gbv,Moxon:2020gha}.
We run CCE on each of the four finite-radius worldtubes that the Cauchy evolution outputs. While in principle the CCEs that use these various worldtubes should yield valid and identical waveforms, we find that because of how the initial data for CCE is constructed there tends to be one worldtube that yields a more physically correct asymptotic waveform. We determine the best worldtube and waveform by examining which waveform's time derivative has the lowest $L^{2}$ norm after the ringdown phase. We find that this test is consistent with the previous method of checking which waveform and Weyl scalars minimally violate the five Bianchi identities~\cite{Mitman:2020bjf,Mitman:2020pbt,Mitman:2021xkq,Mitman:2022kwt}, but tends to yield waveforms with less junk radiation in the inspiral phase. As for the resolution of the CCE, we simply use the highest setting possible which yields errors in the CCE that are well below the errors from the Cauchy evolution. The waveforms that we use are interpolated to a uniform time step of $0.1M$, which is a dense enough time array to capture the important features of the waveform, including those that emerge near and during the merger phase.

Like extrapolated waveforms, CCE waveforms contain `initial data transients'
due to the imperfect initial data on the first null hypersurface of the characteristic evolution. These unphysical features, however, tend to persist much longer than those observed in the extrapolated waveforms. Fortunately, we find that by truncating the earlier parts of 
the CCE waveforms, we can avoid this issue when constructing
our training set in every mode except the $(3,1)$ and $(4,2)$ modes. Therefore, we exclude these modes in our new model. For more details about these modes and why we choose to exclude them, see Appendix~\ref{sec:challenge}.

\subsection{BMS frame fixing}
\label{sec:bms_frame_fixing}

\begin{figure*}[tbh]
	\centering
	\includegraphics[width=\textwidth]{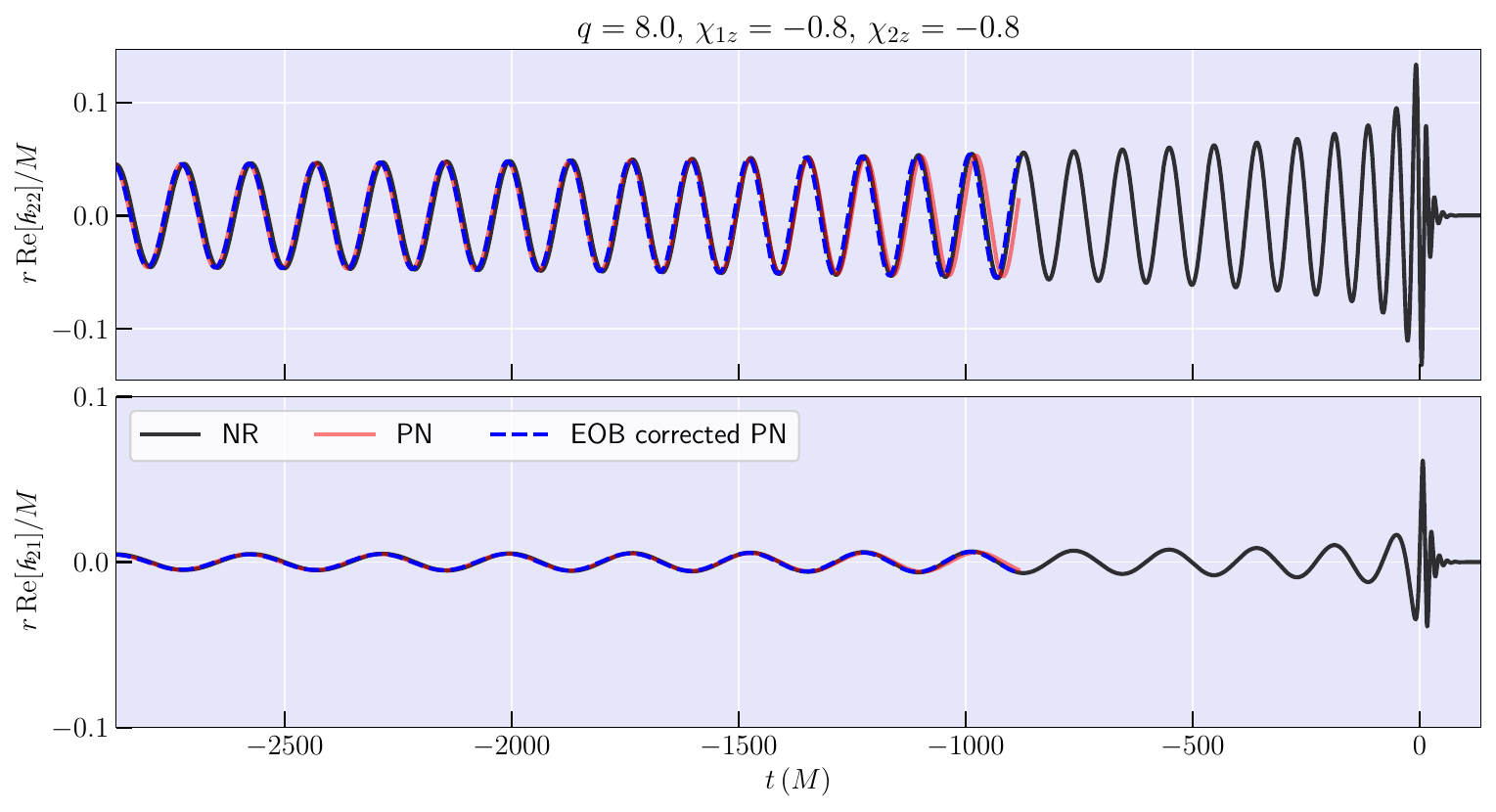}
	\caption{
		NR (black), PN (red), and EOB-corrected PN (blue) waveforms for an example simulation with binary parameters $q=8.0$, $\chiAz=-0.8$, and $\chiBz=-0.8$. The real part of both the $(2,2)$
		and $(2,1)$ modes are shown in the top and bottom panels. Notice that the EOB-corrected PN waveform is more faithful to the NR waveform than the PN waveform.
            }
	\label{fig:pn_eob}
\end{figure*}

When building surrogate models, it is important
to ensure that the training waveforms are 
in the same frame. Otherwise, undesired gauge
artifacts can complicate or even interfere with the various fitting 
and interpolation steps that are used when building the surrogate model. 
The earlier surrogate model, \NRSurOld, for example,
implemented a center-of-mass corrected version of the extrapolated waveforms, in which the waveforms were mapped to a ``Newtonian'' center-of-mass frame of the binary using the coordinate trajectories of the black hole apparent horizons from the simulation~\cite{SXSCatalog}.

However, the true gauge degree of freedom possessed by 
the gravitational waves at future null infinity is not the usual Poincar\'e group, but the
BMS group, which includes an
infinite-dimensional group of transformations that are called 
supertranslations~\cite{Bondi:1962px, Sachs:1962wk} in addition to the usual Poincar\'e transformations. As a result, before using NR waveforms for any analysis, one should
first fix these BMS freedoms.\footnote{Note though that while fixing the BMS frame is important for modeling purposes, e.g., constructing or comparing waveform models,
when examining waveforms at a point on the two-sphere,
the only frame freedom that is relevant is the Poincar\'e
freedom. This is because when looking at a point on the
sky, supertranslations become degenerate with time translations.}
In Refs.~\cite{Mitman:2021xkq,Mitman:2022kwt} this task of fixing
the BMS frame was performed by computing the various BMS charges that
correspond to the symmetries of the BMS group and then finding the
transformations that change those charges in a desired way. For
example, to fix the translation and boost symmetries
Refs.~\cite{Mitman:2021xkq,Mitman:2022kwt} found a transformation
that mapped the center-of-mass charge to have a mean of zero. More specifically, they found the transformation which minimized the time integral of the $L^{2}$ norm of this charge over a three-orbit window. Similarly, for fixing the
system's supertranslation freedom, the same works also found
what supertranslation to apply to the waveforms by examining a
charge known as the Moreschi supermomentum---an extension of the
usual Bondi four-momentum. By finding the supertranslation that
mapped the Moreschi supermomentum to the value expected from PN theory,
it was found that NR waveforms could be made to much better agree with
PN waveforms once they were mapped to the same BMS frame of PN.
This frame is called the PN BMS frame.

Because the surrogate model that we are building is for 
hybrid waveforms in which PN waveforms are stitched to
the NR waveforms, the natural BMS frame to work with is, similarly, the PN BMS frame.
Therefore, to fix the BMS frame of our waveforms we use the frame fixing procedure described in Ref.~\cite{Mitman:2022kwt} and the python module \texttt{scri}~\cite{scri, Boyle:2013nka, Boyle:2014ioa, Boyle:2015nqa}. That is, we fix the translation and boost freedoms by mapping the center-of-mass charge to have a mean of zero, we fix the rotation freedom by mapping the angular velocity vector to be aligned with that of a PN waveform, and we fix the supertranslation freedom by mapping the NR Moreschi supermomentum to agree with the PN Moreschi supermomentum~\cite{Mitman:2022kwt}. We perform this frame fixing using a three-orbit window
that starts $\sim 2500 - 3500M$ before the peak of the $L^{2}$ norm of the NR strain. This choice of BMS frame implies that the $\ell \geq 1$ components of the Moreschi supermomentum of our BBH vanish at $t \rightarrow-\infty$. This is equivalent to matching the 
PN memory terms with the NR system's memory over the hybridization window. This choice also implies that \NRSurNew's strains vanish at $t \rightarrow -\infty$.

\subsection{Hybridization}
\label{sec:hybridization}

Because of computational limits, NR simulations of BBHs typically only cover the last $20$ orbits of inspiral. 
Thus, they are not nearly long 
enough to span the full LVK detection band for stellar mass binaries.
More precisely, the initial frequency of (2,2) mode of these waveforms falls within the LVK band, taken to begin 
at $\flow = 20\textrm{Hz}$, for total masses $M =  m_{1} + m_{2} \gtrsim 60M_{\odot}$. 
To address this limitation and extend the validity of our model
to lower values of total mass, we hybridize the CCE waveforms that are obtained from 
NR with the early inspiral parts of EOB phase-corrected PN waveforms. We create pure PN waveforms 
using the python package \texttt{GWFrames}~\cite{GWFrames}. For the PN orbital phase
we include non-spinning terms up to 4 PN order~\cite{Blanchet:2004ek,Blanchet:2013haa,Jaranowski:2013lca,Bini:2013zaa,Bini:2013rfa} and spinning terms up to 2.5 PN order~\cite{Kidder:1995zr,Will:1996zj,Bohe:2012mr}. For the PN amplitude we include non-memory terms to 3.5 PN order~\cite{Blanchet:2008je,Faye:2012we,Faye:2014fra}, non-spinning memory terms to 3 PN order, and spinning memory terms up to 2 PN order~\cite{Favata:2008yd,Mitman:2022kwt}. 
We use the TaylorT4~\cite{Boyle:2007ft} approximant to compute the PN phase, but we replace this with an EOB-derived phase
for the following reasons.

As noted in the previous work with \NRSurOld~\cite{Varma:2018mmi}, the accuracy of the
inspiral parts from PN waveforms can be improved by replacing the
PN phase with the phase that is derived from an EOB model, which undergoes an NR calibration. This improvement is typically larger
for high mass-ratio systems, where the PN deviation from
NR tends to be more significant, as is shown in Fig.~\ref{fig:pn_eob}. 
For the phase correction\footnote{The phase correction procedure is identical to that described in Sec. IV B of Ref.~\cite{Varma:2018mmi}.} to the PN waveforms used in this new surrogate model, we use the EOB model $\texttt{SEOBNRv4\_opt}$~\cite{Bohe:2016gbl,Devine:2016ovp}.\footnote{While $\texttt{SEOBNRv4\_opt}$ is trained on extrapolated waveforms rather than CCE waveforms, we do not expect this feature to noticeably impact the surrogate because the phase evolution of these different waveforms should still be comparable. Still, it would be interesting to see how our surrogate model construction changes when using EOB waveforms that have been calibrated with CCE waveforms.}

Currently, the waveforms produced by CCE contain `initial data transients'
or `junk radiation' because of imperfect initial data that forces us to discard early parts
of the waveforms~\cite{Mitman:2020pbt}. We find that the transients present in the CCE waveforms typically last noticeably longer
than the junk radiation of the extrapolated waveforms~\cite{Moxon:2021gbv}. Because of this, we instead use a hybridization window that is closer to the merger: roughly $2500 - 3500M$ before the peak of the $L^{2}$ norm of the strain. The later window further 
necessitates our use of the EOB-corrected phase for the inspiral part fo the PN waveforms. 

Once the NR waveforms are mapped to the same frame as the EOB-corrected PN waveforms via the procedures
described in~\ref{sec:bms_frame_fixing},\footnote{Even though we are using EOB-corrected PN waveforms as the target waveform in the BMS frame fixing procedure, because the EOB correction tends to zero as $t\rightarrow-\infty$, the frame that we map our NR waveforms to is still consistent with the PN BMS frame. The EOB-correction simply helps ensure that our mapping to the PN BMS frame is as accurate as possible.} we then hybridize NR and PN together, for each spin-weighted spherical
harmonic mode, $\hlm$, as:
\begin{align}
\hlm^{\text{hyb}}=\hlm^{\text{PN}}+f\left(\frac{t-t_{1}}{t_{2}-t_{1}}\right)\left(\hlm^{\text{NR}}-\hlm^{\text{PN}}\right),
\end{align}
using the following transition function
\begin{align}
\label{eqn:smoothingfunction}
f(x)=
\begin{cases}
	0&x\leq0,\\
	\left(1+\exp\left[\frac{1}{x-1}+\frac{1}{x}\right]\right)^{-1}&0<x<1,\\
	1&x\geq1.
\end{cases}
\end{align}

Our choice of transition function matches that used in Ref.~\cite{Mitman:2021xkq}.
In the previous surrogate, \NRSurOld, instead of hybridizing the complex waveforms directly, the waveforms were decomposed into amplitude and frequency before these components were hybridized independently. This was done to avoid undesirable artifacts that their transition function introduced, which are shown in Fig.~4 of Ref.~\cite{Varma:2018mmi}. 
Because CCE waveforms contain memory, which acts as a time-dependent offset in
the waveforms, the decomposition into amplitude and frequency is not
as useful as it is for the extrapolated waveforms which do not contain
memory. Therefore, we choose to not use the previous hybridization
method and instead directly stitch the NR and PN waveforms
together. We do observe some minor glitches in the orbital frequency of a few of our hybrid waveforms in the hybridization
window. However, these effects are largely negligible when compared to
the other modeling errors for \NRSurNew.

\subsection{Post-processing the training data}
\label{sec:post_processing}

In this section, we now describe how we construct the surrogate model from the hybrid waveforms of Sec.~\ref{sec:hybridization}.

\subsubsection{Down-sampling and common time array}
\label{sec:common_time_array}

First, we apply a time shift to each training waveform such that 
the peak of the $L^{2}$ norm
\begin{equation}
\label{eqn:norm}
A_{\mathrm{tot}}(t)=\sqrt{\sum_{\ell,m}|\hlm(t)|^{2}}
\end{equation}
is aligned at $t=0$. The peak time of this curve is determined from a
quadratic fit using $5$ time samples that are adjacent to the
discrete maximum of $A_{\mathrm{tot}}$. When we compute the sum in Eq.~\eqref{eqn:norm}, we use every mode of the hybrid waveform, including the $m<0$ modes.

To begin with, the length of each hybrid waveform is determined by ensuring that 
the initial orbital frequency is $\omegaorb = 2\times 10^{-4}\,
\mathrm{rad}/M$, where $\omegaorb$ is approximated from $\phi_{22}$, the phase of
the (2,2) mode, using
\begin{equation}
	\omegaorb = \frac{1}{2}\frac{d\phi_{22}}{dt}
\end{equation}
This frequency choice, however, results in
waveforms with different lengths for different mass ratios and spins.
The surrogate-building procedure, however, requires that
every training waveform share a common time array. 
Therefore, to remedy this issue we truncate the training waveforms
such that they begin with time $t=-5.8 \times 10^{8}M$, 
which is the first time of
the shortest waveform in the surrogate's training set.
After truncation, the training set's largest starting orbital
frequency is $\omegaorb = 2.8\times10^{-4} \, \mathrm{rad}/M$. Consequently, this frequency is the low-frequency limit of validity for the surrogate model.

For the LVK observatories, if we assume that $20\mathrm{Hz}$ is 
the lowest GW frequency that can be measured, then the $(2,2)$ mode of the surrogate model can be considered valid for total masses $M \geq 0.9 M_{\odot}$.
The highest spin-weighted spherical harmonic mode included
in \NRSurNew is the $(5,5)$ mode for which the corresponding frequency
is a factor of $5/2$ more than that of the $(2,2)$ mode. Thus, the entire surrogate is valid for total masses $M \geq 2.25 M_{\odot}$.

Because the hybrid waveforms are millions of $M$ long, it is not practical
to sample the entire waveform with a small uniform time step like
$0.1M$, as is typically used for NR-only surrogates~\cite{Varma:2019csw}.
Fortunately, the early inspiral, low-frequency portion of the waveform does not 
require as dense a sampling as the later high-frequency portion. Therefore, we instead
down-sample the time arrays of the truncated waveforms such that there are
only 5 points per orbit for the shortest hybrid waveform of the training set.
However, for $t\geq -1000M$, we switch to uniformly spaced time samples
with a time step of $0.1M$ to ensure that we have sufficiently dense sampling
for the late inspiral and merger-ringdown phases where the frequency reaches its peak.
We retain times up to $135M$ after the peak to ensure that we fully capture the numerically resolvable parts of the
ringdown phase.

Given the common down-sampled time array, we then use cubic splines
to interpolate all of the waveforms in the training set onto the common time array. However, we first
transform the waveforms to the co-orbital frame, which we construct via
\begin{equation}\label{eqn:coorb_decomp}
\text{co-orbital frame: }\left\{
\begin{aligned}
&\hlm^{C} = \hlm e^{i m\phi_{\mathrm{orb}}} \\
&\h_{22} = A_{22} e^{i\phi_{22}} \\
&\phi_{\mathrm{orb}} = \phi_{22}/2 
\end{aligned}
\right .
\end{equation}
where $\hlm$ is the inertial frame waveform, $\phi_{\mathrm{orb}}$
is the orbital phase, and $A_{22}$ and $\phi_{22}$ are the amplitude
and phase of the $(2,2)$ mode. The co-orbital frame is
roughly co-rotating with the binary and is obtained by
applying a time-dependent rotation about the $\hat{z}$-axis by an 
amount measured by the instantaneous orbital phase. As 
the waveforms are slowly varying functions of time in the co-orbital frame,
by transforming to this frame we can increase the interpolation accuracy.
For the $(2,2)$ mode, we sample the amplitude $A_{22}$ and phase $\phi_{22}$,
while for all other modes we use the real and imaginary parts of $\hlm^{C}$.

\subsubsection{Phase alignment}
\label{sec:phase_alignment}

After interpolating to a common time array, we align the phases of the waveforms by rotating the waveforms
about the $\hat{z}$-axis such that the orbital phase $\phi_{\mathrm{orb}}$ is zero at time $t=-1000M$. This ensures that each waveform corresponds to a binary with its heavier black hole on the $\hat{x}$-axis at that time.
Note that this frame is constructed using information from the waveform at future null infinity, and as a result
these BH positions need not correspond to the 
gauge-dependent BH positions in the NR simulations. 

\subsubsection{Data decomposition}
\label{sec:data_decomposition}

As mentioned earlier, it is easier to build a model for slowly varying functions of time. 
Because of this, we decompose the inertial frame strain $\hlm$, which is oscillatory, into 
simpler waveform data pieces and build a separate surrogate model for each
of these data pieces. 
For the $(2,2)$ mode, we decompose this mode into the amplitude $A_{22}$ 
and the phase $\phi_{22}$, while for the other $m \neq 0$ waveform modes, we 
model the real and imaginary parts of the co-orbital frame strain, 
$\h_{lm}^{C}$, using Eq.~\eqref{eqn:coorb_decomp}.
For the $m = 0$ modes 
of non-precessing systems, $\hlm^{C}$ is purely real (imaginary) for even (odd) $\ell$. Because of this, we only model the non-trivial part for the $m = 0$ modes. 

Because our hybrid waveforms are rather long---extending over
roughly $3\times 10^{4}$ orbits---$\phi_{22}$
roughly spans $10^{5}$ radians.
Accurately modeling the phase evolution of such long hybrid waveforms
poses a challenge.
We find that we can resolve this issue, however, by subtracting the leading-order Taylor T3 PN phase~\cite{Damour:2000zb}, $\phi_{22}^{\mathrm{T3}}$,
and simply modeling the phase residual, $\phi_{22}^{\mathrm{res}}=\phi_{22}-\phi_{22}^{\mathrm{T3}}$,
as was performed in Ref.~\cite{Varma:2018mmi}. The leading-order prediction from the TaylorT3 PN 
approximant~\cite{Damour:2000zb} is
\begin{equation}
\label{eqn:taylor_T3}
\phi_{22}^{\mathrm{T3}} = \phi_{\mathrm{ref}}^{\mathrm{T3}} - \frac{2}{\eta \theta^{5}}
\end{equation}
with
\begin{equation}
\theta = [ \eta (t_{\mathrm{ref}} -t ) / (5M) ]^{-1/8},
\end{equation}
where $\phi_{\mathrm{ref}}^{\mathrm{T3}}$ is an arbitrary integration constant, 
$t_{\mathrm{ref}}$ is an arbitrary time offset,
and $\eta = q/(1+q)^{2}$ is the symmetric mass ratio. Because, by definition, $\theta$
diverges at $t = t_{\mathrm{ref}}$, to avoid such divergences we set $t_{\mathrm{ref}} = 1000M$, which is sufficiently long 
after the end of the waveform. We also choose $ \phi_{\mathrm{ref}}^{\mathrm{T3}}$
such that $\phi_{22}^{\mathrm{T3}} = 0$ at $t=-1000M$, i.e., the time at which we align the phase, as outlined in
Sec.~\ref{sec:phase_alignment}.
 
\subsection{Surrogate building}
\label{sec:surrogate_building}

Given the decomposed waveform data pieces, we build a surrogate
model for each individual data piece using the same procedure as that of Sec.~V. C of Ref. \cite{Varma:2018mmi}, with a few minor modifications, which we summarize below.

For each waveform data piece, we begin by constructing a linear basis in parameter space, so that we can reduce the training data set to a smaller representative data set. The basis functions that we use are chosen in the following iterative manner~\cite{Field:2011mf,Field:2013cfa,binev2011convergence,devore2013greedy}, called the `greedy algorithm':
\begin{enumerate}
\item Pick out the training data with the largest $L^{2}$ norm and add it to the basis set as the first basis function;
\item Compute the projection error between each of the training data and the basis set;
\item Determine which of the training data has the highest projection error and add this to the basis set;
\item Repeat steps 2--3 until a pre-determined number of basis functions for each data piece is obtained.
\end{enumerate}
For step 4, we determine the number of basis functions used for each of the data pieces through trial and error. That is, we increase the number of basis functions until the inclusion of new basis functions introduces noise into the model or gives diminishing returns in terms
of minimizing the projection errors. The number of basis functions that is used for each data piece is shown in Table~\ref{tab:basis_number}.
\begin{table}[h!]
\label{tab:basis_number}
\centering
\begin{tabular}{@{}l@{\hspace*{7mm}}c@{\hspace*{7mm}}c@{}}
	\Xhline{3\arrayrulewidth}
  data piece & number of basis functions\\[5pt]
  $A_{22},\,\phi_{22}$ & 15 \\
  $\text{Re}\,\h^{C}_{21},\, \text{Im}\,\h^{C}_{21}$ & 12 \\
  $\text{Re}\,\h^{C}_{20}$ & 12 \\
  $\text{Im}\,\h^{C}_{30}$ & 12 \\  
    $\text{Re}\,\h^{C}_{43},\, \text{Im}\,\h^{C}_{43}$ & 12 \\
   $\text{Re}\,\h^{C}_{55},\, \text{Im}\,\h^{C}_{55}$ & 8 \\
  others & 10 \\
	\Xhline{3\arrayrulewidth}
\end{tabular}
\caption{Number of basis functions for each data piece.}
\end{table}

Next, we build empirical time interpolants~\cite{chaturantabut2010nonlinear,Maday:2009,Hesthaven:2014,antil2013two,Field:2013cfa} 
with the same number of empirical nodes as the number of basis functions
that are used to model the data piece. Following the methodology of Ref.~\cite{Varma:2018mmi} we also require that the start of the waveform always be
included as one of the empirical nodes. This provides
an `anchor point' that ensures that the waveform data pieces start with the correct value. In Ref.~\cite{Varma:2018mmi}, 
no empirical nodes were picked at times past $t>50M$
to ensure that little to no numerical noise was being modeled, particularly for
the phase data piece. We follow the same convention
for the phase data pieces; however, for the other data pieces we allow time nodes past
$t>50M$ to ensure
that the surrogate correctly models the memory throughout the entirety of the ringdown phase.

Finally, for each empirical time node, we construct a
parametric fit for the waveform data piece, following the Gaussian
process regression (GPR) fitting method described in the supplementary
material of Ref.~\cite{Varma:2018aht}, using the
python package \texttt{scikit-learn}~\cite{scikit}. For the fit itself, we use the parameterization used in Ref.~\cite{Varma:2018mmi}. That is, we use the parameters $\log(q), \, \hat{\chi},$ and $\chi_{a}$, where $\hat{\chi}$
is the leading order spin parameter~\cite{Khan:2015jqa,Ajith:2011ec,Cutler:1994ys,Poisson:1995ef} for the GW phase in the PN expression
\begin{align}
	\hat{\chi}=\frac{\chieff - 38 \eta ( \chiAz +\chiBz )/113}{1-76\eta/113} ,
\end{align}
with
\begin{align}
	\chieff=\frac{q \chiAz+ \chiBz}{1+q},
\end{align}
and $\chi_{a}$ is the anti-symmetric spin defined as
\begin{align}
\chi_{a}=\frac{1}{2}(\chiAz-\chiBz).
\end{align}

\begin{figure}[!ht]
	\centering
	\includegraphics[width=0.5\textwidth]{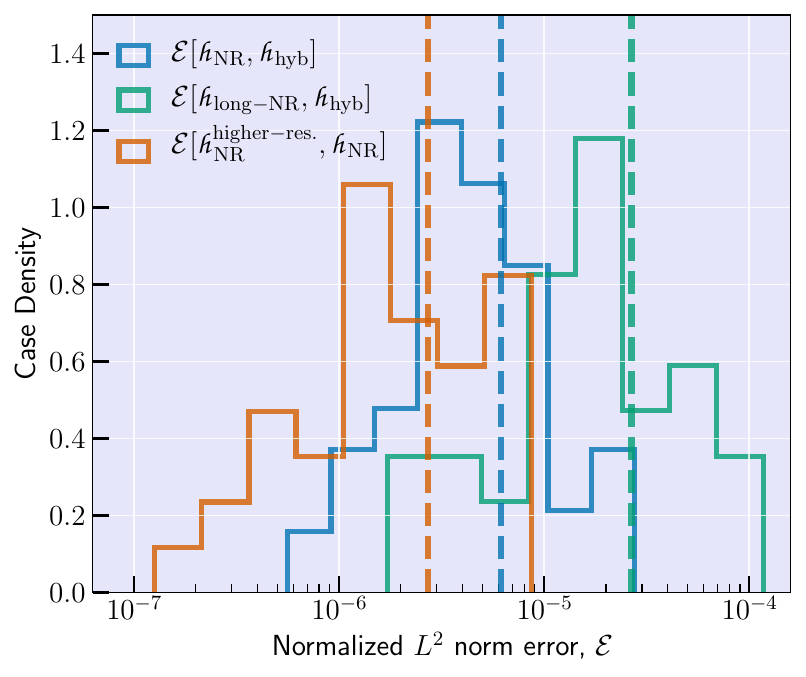}
	\caption{Two unique estimates of the hybridization error.
		$\mathcal{E}\left[\h_{\mathrm{NR}},\h_{\mathrm{hyb}}\right]$, using Eq.~\eqref{eqn:errors}, computes the error between the hybrid and the NR waveforms in the hybridization window. Meanwhile, $\mathcal{E}\left[\h_{\mathrm{long-NR}},\h_{\mathrm{hyb}}\right]$ computes the error
		between the hybrid and the long NR waveforms from an initial time of $t_{1}=-6000M$ to
		the end of hybridization window. We also include 
		 $\mathcal{E}\left[\h^{\mathrm{higher-res}}_{\mathrm{NR}},\h_{\mathrm{NR}}\right]$ as a resolution error between the
		 two highest resolution NR waveforms, computed within the hybridization window.
		$\mathcal{E}\left[\h_{\mathrm{NR}},\h_{\mathrm{hyb}}\right]$
		was computed for every one of the $102$ training waveforms, while $\mathcal{E}\left[\h_{\mathrm{long-NR}},\h_{\mathrm{hyb}}\right]$ and $\mathcal{E}\left[\h^{\mathrm{higher-res}}_{\mathrm{NR}},\h_{\mathrm{NR}}\right]$ were computed for the $37$ waveforms for which longer and higher resolution simulations were available.
		The dashed lines represent the median values.}
	\label{fig:hyb_err}
\end{figure}

\section{Error Quantification}
\label{sec:errors}

With the methodology behind the construction of our surrogate model outlined in Sec.~\ref{sec:methods}, we now examine the quality of \NRSurNew by conducting a variety of model consistency checks and waveform comparisons. This process involves examining both the error in the hybridization and the error in the surrogate model itself as well as the ability of the surrogate to model waveforms outside of its training parameter space.

\begin{figure*}[t!]
	\centering
	\includegraphics[width=\textwidth]{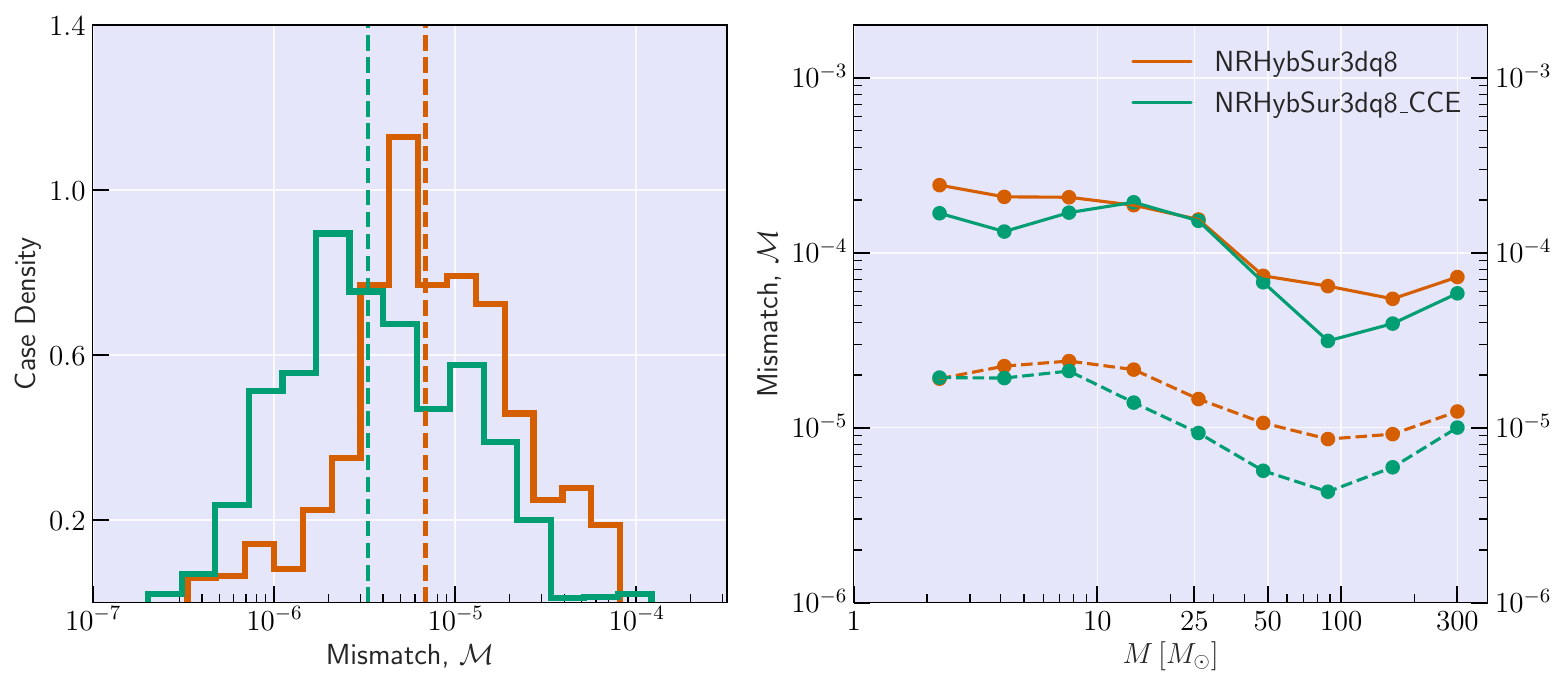}
\caption{
Frequency-domain mismatches for the \NRSurOld and \NRSurNew surrogates when
compared to their respective training hybridized waveforms (extrapolated and CCE,
respectively). The results for \NRSurOld are taken from Fig.~6 of Ref.~\cite{Varma:2018mmi} and are plotted for comparison with \NRSurNew. For both of these surrogate models, we compute leave-five-out errors at
several points in the sky of the source frame using all available modes in each
model: $\ell \leq 4$ and $(5,5)$, but excluding $(3,1)$, $(4,2)$, and $(4,1)$ for \NRSurNew and $(4,1)$ and
$(4,0)$ for \NRSurOld. \bfemph{Left}: Mismatches computed using a flat noise curve
for only the late inspiral part (NR part) of the hybrid waveforms. 
The dashed vertical line represents
the median mismatch value. \bfemph{Right}: Mismatches computed using the Advanced-LIGO noise curve as a function of the system's total mass. The solid (dashed) line represents
$95^{\mathrm{th}}$ percentile (median) mismatch values.
}
\label{fig:waveform_mismatch}
\end{figure*}

\subsection{Hybridization errors}
\label{sec:hyberr}

Before building the surrogate model, we first have to map the NR waveforms to be in the same frame as the EOB-corrected PN waveforms. As described in Sec.~\ref{sec:bms_frame_fixing}, this fixes the BMS freedom of the NR waveforms by mapping the waveforms to the PN BMS frame using the procedure outlined in Ref.~\cite{Mitman:2022kwt}. Following this, we then stitch the two waveforms to create
a hybrid waveform that we use as our training waveform. One of the important checks for the quality of our final training waveform is to understand the errors that result from performing this hybridization procedure between NR and PN waveforms. There are two natural ways to conduct this check.

First, we can simply compare the
hybrid waveform to the NR waveform in the hybridization window. 
This check will determine the combined discrepancy that results from the discrepancy 
between the PN and NR waveforms and from using the smooth transition function in Eq.~\eqref{eqn:smoothingfunction}.
Next, to examine the error introduced from using PN for the early inspiral parts
of our hybrid waveform, we can compare the hybrid to a
simulation that has identical binary parameters, but a larger initial BH
separation. Fortunately, of the simulations that we include in our surrogate's training data set, 
37 of them have these longer extensions that we can
use for a comparison with the hybrid waveforms. 

To quantify the error for these comparisons, we use the following function: 
\begin{equation}
\label{eqn:errors}
\mathcal{E}\left[\h^{(1)}, \h^{(2)}\right]\equiv\frac{1}{2} \frac{\sum\limits_{\ell,m} 
	\int_{t_{1}}^{t_{2}} \lvert \hlm^{(1)}(t) - \hlm^{(2)} (t)\rvert^{2} dt}{\sum\limits_{\ell,m} \int_{t_{1}}^{t_{2}}\lvert \hlm^{(1)}(t) \rvert^{2} dt}.
\end{equation}

For the first comparison, we compute the error within the hybridization window, where $t_{1}$ is roughly $3000M$ before merger and $t_{2}$ is the time at which the system has undergone $3$ orbits since $t_{1}$.
For the latter comparison, we
first map the long NR waveforms to the BMS frame of the training NR waveforms
and then compute the error between these waveforms from $t_{1}=-6000M$ to the
end of the hybridization widow. The results of these comparisons are shown in
Fig.~\ref{fig:hyb_err}. We observe that our two estimates for the hybridization error
are low, but tend to be higher than the estimate of the NR resolution error.
Note, though, that the NR resolution errors, which were computed for a smaller
subset of NR simulations for which the higher resolution data was available, come from simulations that are for more comparable mass binaries. Because the difficulty
of NR simulations increases with mass ratio, we therefore expect that the NR resolution
error that we computed is a minor underestimate of the true error. Regardless, Fig.~\ref{fig:hyb_err} suggests that our training waveforms are primarily limited by the accuracy of the PN waveforms, rather than the NR resolution error.

\subsection{Model errors}
\label{sec:surrogate_errors}

We now evaluate the accuracy of the surrogate model
\NRSurNew by comparing the waveforms that it produces to the hybridized PN/NR waveforms that were used to train it.
We quantify this model accuracy by computing the frequency-domain mismatch $\mathcal{M}$ between two waveforms $\h_{1}$ and $\h_{2}$ via
\begin{align}
\mathcal{M} &= 1 - \frac{\braket{\h_1,\h_2}}{\sqrt{\braket{\h_1,\h_1}\braket{\h_2,\h_2}}}
\label{eqn:mismatch}
\end{align}
with
\begin{align}
\braket{\h_1,\h_2} &= 4 \text{Re} \left[ \int_{f_{\mathrm{min}}}^{f_{\mathrm{max}}} \frac{\tilde{\h}_1(f)\tilde{\h}^*_2(f)}{S_n(f)}\,df\right],
\label{eqn:inner_fd}
\end{align}
where $\tilde{\h}(f)$ denotes the Fourier transform of the strain $\h(t)$, 
$^{*}$ the complex conjugate, and
$S_{n}(f)$ the one-sided power spectral density of, say, a GW detector. The mismatches are optimized over shifts in time, 
polarization angle, and the initial orbital phase following the procedure described in Appendix D of Ref.~\cite{Blackman:2017dfb}. For each pair of waveforms, we compute mismatches at a total of 37 sky points that are uniformly distributed over the two-sphere. 

Before performing the Fourier transform, we taper both ends of the time domain waveform.\footnote{A value mismatch at both ends of the waveform tends to result in the presence of Gibbs phenomenon in the Fourier spectrum. To avoid this, we taper the waveform to zero at both ends using a Planck window~\cite{McKechan:2010kp}.} The tapering at the start of the waveform is done over $1.5$ cycles of the $(2,2)$ mode.
Because the waveforms that we are examining include memory effects, tapering them in the ringdown region
can produce a significant level of windowing effects\footnote{A new scheme for pre-processing that potentially reduces this windowing effect is proposed in Ref.~\cite{EthPIP}.} in the Fourier spectrum. Therefore, before computing mismatches we first pad the end of the waveforms with their final values for $1000M$ and then taper them over this padded region.
We find that a padding length of $1000M$ is enough to significantly reduce windowing effects from tapering
and that the mismatch result is not very sensitive to this choice of padding length.

Because we use all available hybrid waveforms for the training of the model \NRSurNew, if
we compute the mismatch of our model against hybrid waveforms we would obtain
a training error, rather than an estimate of the true modeling error. Thus, we instead
estimate the out-of-sample error by performing leave-five-out analyses for the \NRSurOld and \NRSurNew models.
We construct exactly $20$ trial surrogates, leaving out $5$ or $6$
waveforms from the training set for each surrogate. By calculating the mismatch
between each surrogate and the left-out waveforms, we are then able to assess the 
performance of each surrogate against waveforms that were not used in the training process.

\begin{figure}[t]
\centering
\includegraphics[width=0.5\textwidth]{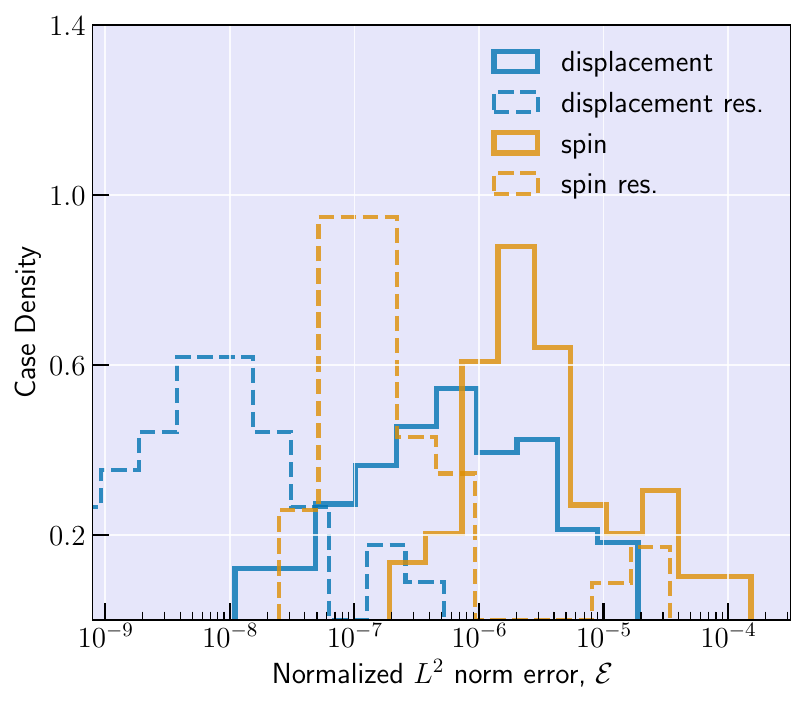}
\caption{Errors for the displacement (blue) and spin (orange) memory effects computed by comparing the training waveforms to \NRSurNew's output using Eq.~\eqref{eqn:errors}. For this result, we show the leave-five-out cross-validation errors (solid line). As a reference, we also include the errors computed between the two highest resolution waveforms (dashed line).}
\label{fig:memory_mis}
\end{figure}

\begin{figure}[t]
\centering
\includegraphics[width=0.497\textwidth]{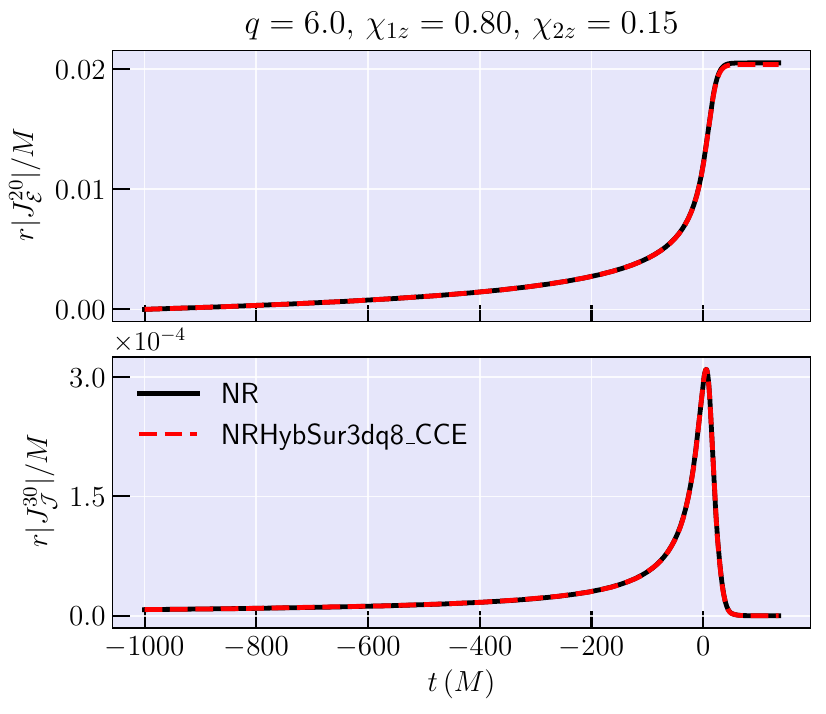}
\caption{Amplitudes of the most dominant modes of the displacement and spin memory effects when computed from the NR waveform and the surrogate evaluation for the case with the largest errors in Fig.~\ref{fig:memory_mis}. The top and bottom panels show the $(2,0)$ mode of the displacement null memory and the $(3,0)$ mode of the spin null memory.}
\label{fig:mem_worstcase}
\end{figure}

The left panel of Fig.~\ref{fig:waveform_mismatch} 
shows the mismatches for \NRSurNew that are computed using a flat (white) noise curve $(S_{n}=1)$ over the
late inspiral part (NR part) of the hybrid waveforms, truncating
the waveforms to start at $t=-3500M$ for \NRSurOld and at $t=-2500M$ for
\NRSurNew.\footnote{The discrepancy is due to the later hybridization window that was used
for \NRSurNew as explained in Sec.~\ref{sec:hybridization}. We define 
$f_{\mathrm{min}}$ to be the frequency of the $(2,2)$ mode at the end of the initial tapering window, and $f_{\mathrm{max}} = 5f^{\mathrm{peak}}_{22}$, where
$f^{\mathrm{peak}}_{22}$ is the frequency of the $(2,2)$ mode at its peak. The choice of $f_{\mathrm{max}}$ ensures that we capture
the peak frequency of every mode.}

The right panel of Fig.~\ref{fig:waveform_mismatch} show results that are more relevant to GW observations: namely, the mismatches computed with the 
Advanced-LIGO design sensitivity Zero-Detuned-High-Power noise curve~\cite{aLIGODesignNoiseCurve} with $f_{\mathrm{min}}=20\mathrm{Hz}$ and $f_{\mathrm{max}}=2000\mathrm{Hz}$ for various total masses.
We also include the mismatch result for \NRSurOld from Ref.~\cite{Varma:2018mmi} for comparison. 
On the horizontal axis, different total masses correspond to different portions of the waveform falling within the $[20\mathrm{Hz},2000\mathrm{Hz}]$ window. This roughly implies that the low end of the total mass axis is a proxy for the fidelity of the early inspiral part of the waveforms,
while the high end is a proxy for the late inspiral parts. We show the mismatch for various total
masses: from the lower limit of the range of validity of the surrogate, i.e.,
$M \gtrsim 2.25M_{\odot}$, up to $M=300M_{\odot}$. For each total mass point that we plot, we show both the median and the $95^{\mathrm{th}}$
percentile mismatches.

For the surrogate modeling errors, we obtain values that are comparable to---and often better than---those of \NRSurOld, despite the additional 
modeling challenges resulting from the new contributions due to the presence of memory. 
The $95^{\mathrm{th}}$ percentile mismatches fall below $\sim 3 \times 10^{-4}$ for the
entire mass range.

Last, we test how well the surrogate models both the displacement and spin memory contributions. To do this, we compute
the displacement and spin null memories, i.e., Eqs. (17b) and (17d) from Ref.~\cite{Mitman:2020bjf} which only depend on the strain, using both
the training waveforms and the surrogate evaluation, and then calculate the normalized $L^{2}$ norm error between the two using Eq.~\eqref{eqn:errors} 
over the $t\in[-1000M, 135M]$ window. The choice of the window's starting time
is arbitrary but early enough to capture most of the memory.
For this check, we again perform leave-five-out 
analyses to estimate \NRSurNew's modeling error.
As a reference for these errors, we also show the numerical resolution error from the same set of 37 simulations used for Fig.~\ref{fig:hyb_err}.
As shown in Fig.~\ref{fig:memory_mis}, the modeling error
for the two memory effects is at a reasonable level, albeit higher than our estimate of the resolution error. As shown in Fig.~\ref{fig:mem_worstcase}, even for the case that corresponds to the largest error value in Fig.~\ref{fig:memory_mis}, the memory effects computed from the surrogate evaluation closely agree with those computed from NR waveforms.

We suspect that the spin memory accuracy is typically worse than that of the displacement memory because the spin memory is smaller than the displacement memory and is thus harder to resolve. The discrepancy between the modeling error and the resolution error suggests that
there is room for future improvements in modeling the memory contribution. Building future surrogate models using CCE waveforms from higher resolution and longer NR simulations might help improve this modeling error; however, a new data decomposition scheme or even a new modeling strategy could be necessary to obtain an improved modeling of memory effects.

\subsection{Extrapolating outside training region}
\label{sec:extrapolating}

The errors that we have examined thus far have been restricted to the training 
region of the parameter space: $q\leq 8, \, \chiAz, \, \chiBz 
\in [-0.8, 0.8]$. It is possible, however, to evaluate
the surrogate outside the training region, e.g., for larger mass ratio, $q$, or even
higher primary or secondary
spin magnitudes: $\lvert \chiAz \rvert, \, \lvert \chiBz \rvert$. 
Consequently, to understand the extrapolation capability of the model, we compute
errors of the model against a few existing simulations (SXS:BBH:0185, 0189, 0199, 1124
2085, 2105, 2132, and 2515)~\cite{SXSCatalog,Boyle:2019kee,Blackman:2015pia,Chu:2015kft} that have relatively high mass ratios
or spin magnitudes. 

\begin{figure}[t]
	\centering
	\includegraphics[width=0.5\textwidth]{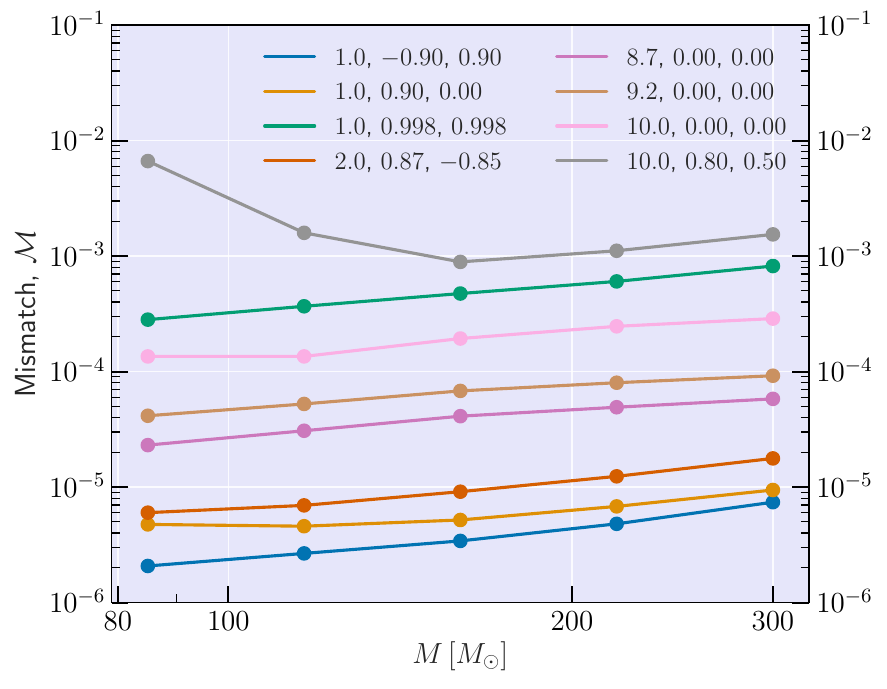}
	\caption{Noise-weighted frequency-domain mismatches between \NRSurNew and
		NR simulations that are outside the model's training region. The numbers in the legend correspond to the NR simulation's mass ratio $q$, primary spin $\chiAz$, and secondary spin $\chiBz$. The mismatches that are shown are
		computed at several points in the sky of the source frame using the Advanced-LIGO 
		noise curve. Each of the solid
		lines represents the $95^{\mathrm{th}}$ percentile mismatch values.}
	\label{fig:ligomm_ext}
\end{figure}

As shown in Fig.~\ref{fig:ligomm_ext}, the mismatch results, while worse than those shown in Fig.~\ref{fig:waveform_mismatch}, are nonetheless reasonable. The highest three mismatch results correspond to 
the three most extreme parameter simulations: mass ratio $q=10$ or spin magnitude $|\chiAz|,|\chiBz|=0.998$, for which we do
not expect the \NRSurNew to perform well. Apart from these, we find that the surrogate
performs well for a modest degree of extrapolation,
with many of the mismatches falling below values near $\sim10^{-4}$. 

\begin{figure}[t]
	\centering
	\includegraphics[width=0.5\textwidth]{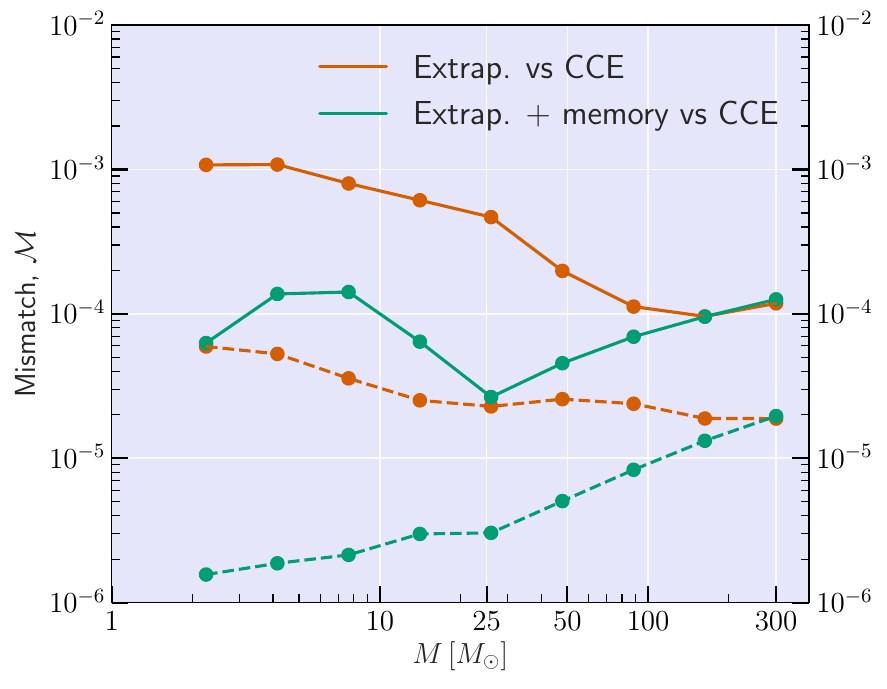}
	\caption{Noise-weighted frequency-domain mismatches between extrapolated and
	CCE hybrid waveforms in red. These are the hybrid waveforms that were used to train \NRSurOld and \NRSurNew. Apart from this, we also include the mismatches between the memory-corrected extrapolated
		and CCE waveforms in green. The mismatches that are shown are computed at several points in the sky of the source frame using the Advanced-LIGO noise curve. The solid (dashed) 
		line represents the $95^{\mathrm{th}}$ percentile (median) mismatch values.}
	\label{fig:ligomm_CCEvEXT}
\end{figure}

\subsection{Systematic bias in waveforms that omit memory or add it through post-processing}
\label{sec:oldnew_compare}

Finally, in Fig.~\ref{fig:ligomm_CCEvEXT}, we show the mismatch between the
training waveforms used for \NRSurNew and those that were used for \NRSurOld. The primary purpose of
this analysis is to obtain a rough estimate of the level of systematic bias that one
could expect from analyzing a GW signal that contains memory using a waveform
that does not. Apart from this, we also show the mismatch between an extrapolated waveform, once we have added the expected memory contribution to it using Eq. (17b) of
Ref.~\cite{Mitman:2020bjf}, and a CCE waveform from the same simulation. This
highlights that, while there is a noticeable difference between CCE waveforms
and extrapolated waveforms, these discrepancies are largely reduced
by adding memory to the extrapolated waveform. Because of this result, we suspect that
memory-detection studies that have used this memory-correction technique, like
Refs.~\cite{Hubner:2019sly,Boersma:2020gxx,islam2021survey,Grant:2022bla,Gasparotto:2023fcg},
would likely obtain similar estimates had they used our new surrogate model
\NRSurNew. Note though that \NRSurNew will be better for performing analyses of the spin memory, seeing as the contribution of spin memory to extrapolated waveforms cannot be as easily corrected~\cite{Mitman:2020bjf}. Regardless, it would still be interesting to see if the
conclusions made by these studies on memory, or even parameter estimation results, change when using \NRSurNew
instead of \NRSurOld.

\section{Conclusion}
\label{sec:conclusion}
In this work, we present a new surrogate model, \NRSurNew, the first
GW model to contain both the oscillatory and memory components of the strain.
Consequently, \NRSurNew is the first model to fully capture
the expected GW physics of binary black hole mergers.
The model is trained on 102
NR/PN hybrid waveforms from aligned-spin binary BH systems with mass ratios
$q\leq 8$ and aligned spins $|\chiAz|,\,|\chiBz|\leq0.8$. 
These hybrid waveforms are constructed by first mapping CCE waveforms to the PN BMS frame before hybridizing them with PN waveforms whose phase has been corrected using EOB waveforms.
Performing this frame fixing helps eliminate unwanted gauge artifacts that could 
potentially interfere with modeling.
The model includes
$\ell \leq 4$ and $(5,5)$ spin-weighted spherical harmonic modes, but not the $(3,1)$, $(4,2)$, or $(4,1)$ modes, and spans the
entire LVK band (with $\flow=20\text{Hz}$) for total masses $M\geq
2.25 M_{\odot}$. By conducting a series of leave-five-out cross-validation analyses, we find
that \NRSurNew can accurately reproduce the hybrid waveforms that were used to train it
with mismatches below $\sim 3\times10^{-4}$ for total masses in the range
$2.25M_{\odot} \leq M \leq 300M_{\odot}$. 
These accuracies are on
par with---and often better than---the previous aligned-spin NR surrogate model, \NRSurOld,
despite the modeling challenges that result from the inclusion of new
modes and memory effects. Apart from this, we also importantly find that \NRSurNew can
successfully capture the null memory contributions with mismatches below
$\sim2\times10^{-4}$. Last, \NRSurNew is also found to reproduce waveforms outside of its trained region of parameter space
for a moderate degree of extrapolation; however, we advise caution when extrapolating
the model. This new model is made publicly available through the python
package \texttt{gwsurrogate}~\cite{gwsurrogate}.

With the expected advances in detector sensitivity for both current and future
gravitational wave observatories, waveform templates with memory effects will
prove to be crucial for analyzing future compact binary detections. The new
surrogate model $\NRSurNew$ serves as the first step in an important endeavor
to produce a complete set of waveform templates that contain these undetected effects and
thus correctly capture the expected gravitational wave physics of binary black hole mergers.

\begin{acknowledgments}

This work was supported in part by the Sherman
Fairchild Foundation and by NSF Grants PHY-2011961, PHY-2011968, and
OAC-2209655 at Caltech, and NSF Grants PHY-2207342 and OAC-2209655
at Cornell. \texttt{SpECTRE} uses \texttt{Charm++}/\texttt{Converse}~\cite{laxmikant_kale_2020_3972617}, which was developed by the
Parallel Programming Laboratory in the Department of Computer Science at the
University of Illinois at Urbana-Champaign.
V.V. acknowledges support from the European Union's Horizon 2020 research and
innovation program under the Marie Skłodowska-Curie grant agreement No.~896869.
L.C.S. was partially supported by NSF CAREER Award PHY-2047382.
S.E.F acknowledges partial support from NSF Grant PHY-2110496 and 
by UMass Dartmouth's Marine and Undersea Technology (MUST) Research
Program funded by the Office of Naval Research (ONR) 
under Grant No. N00014-23-1–2141.
This material is based upon work supported by NSF's LIGO Laboratory which is a
major facility fully funded by the NSF.
This project made
use of \texttt{Python} libraries including \texttt{SciPy}~\cite{2020SciPy-NMeth} and \texttt{NumPy}~\cite{harris2020array}, and 
the figures were produced using \texttt{matplotlib}~\cite{Hunter:2007} and \texttt{seaborn}~\cite{Waskom2021}.
\end{acknowledgments}


\appendix

\section{Challenges in modeling certain modes} 
\label{sec:challenge}

\subsection{Impact of long-lived transient junk}

The waveforms produced by CCE 
contain initial data transients that typically persist much longer than those of the Cauchy evolution.
Because of this effect, as was outlined in Sec.~\ref{sec:hybridization}, we choose to remove the early parts of our NR waveforms when constructing our surrogate's training waveforms.
Nevertheless, even after we perform this truncation, some of the waveforms that we use for the surrogate still exhibit unphysical effects that we do not see when using extrapolated waveforms, or CCE waveforms from higher resolution or longer Cauchy evolutions. 

\begin{figure}[h]
	\centering
	\includegraphics[width=0.5\textwidth]{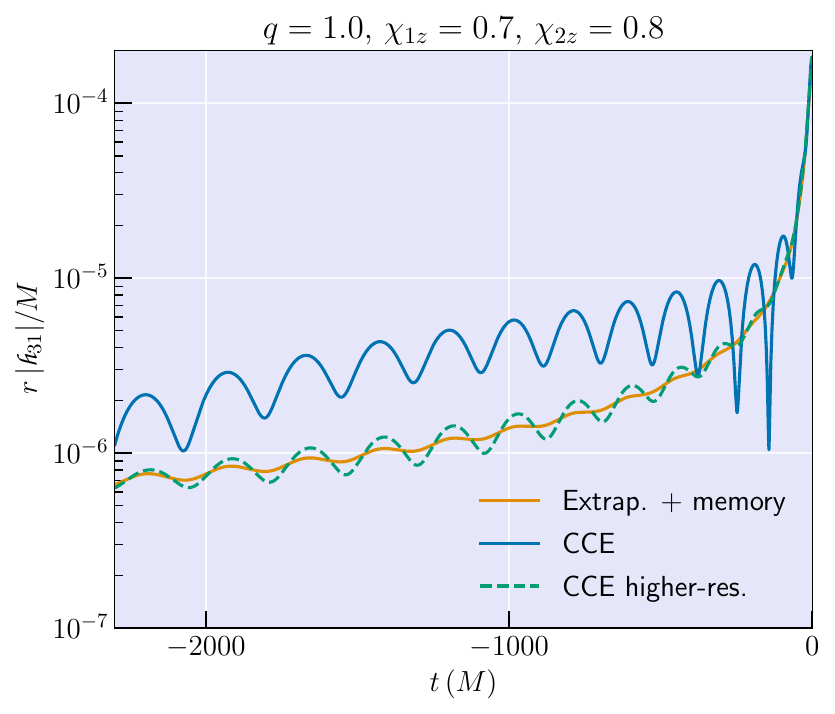}
	\caption{An example of the unphysical oscillations that are seen in the amplitude of the $(3,1)$ mode for one of the CCE training waveforms (blue). As a reference, we also show the extrapolated waveform (orange) for the same simulation and the higher-resolution CCE waveform (green). 
            }
	\label{fig:osc_plot}
\end{figure}

\begin{figure*}[t]
	\centering
	\includegraphics[width=1.0\textwidth]{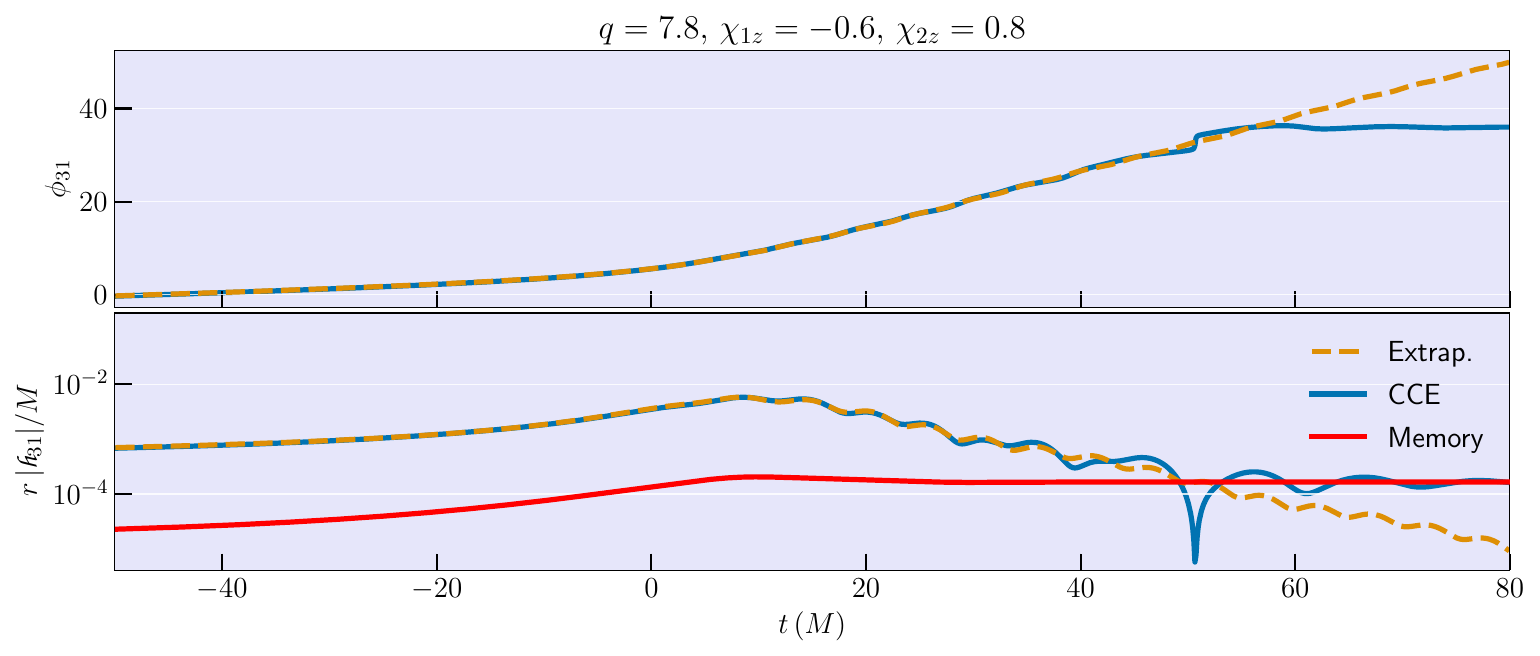}
        \caption{Phase and amplitude of the $(3,1)$ mode for a CCE (blue) and an extrapolated (orange) waveform. We also include the memory contribution (red) to highlight the fact that the CCE waveform does not decay to zero as $t\rightarrow\infty$.}
	\label{fig:rd_plot}
\end{figure*} 

We find that initial data transients are most pronounced in the $(3,1)$ and $(4,2)$ modes, and are often identified by unphysical amplitude oscillations, as
shown in Fig.~\ref{fig:osc_plot}. 
As is the case with all models, the surrogate model \NRSurNew is only as good as the data that it is trained on.
In fact, we find that having even only a few training waveforms with these issues can result in 
noisy and unphysical features being modeled by the surrogate.

We find that these unphysical oscillations in certain modes are
significantly reduced for waveforms extracted from both higher resolution and longer Cauchy evolutions. Unfortunately, such simulations only 
exist for one third of our training data set. Therefore, we instead choose to not model
the $(3,1)$ and $(4,2)$ modes in our new surrogate \NRSurNew. We also omit the $(4,1)$ mode because it is subdominant to the $(4,2)$ mode and, as a result, does not significantly impact the overall waveform accuracy provided that the $(4,2)$ is already excluded. For future surrogate models that are built using CCE waveforms, it is important that we have higher resolution and longer NR simulations to avoid these issues until the problem of constructing initial data for CCE is resolved.

\subsection{Impact of a simulation-dependent BMS frame}

As stressed in Sec.~\ref{sec:bms_frame_fixing}, it is crucial to ensure that the training waveforms 
are in the same BMS frame to avoid undesired frame artifacts that tend to complicate
the waveform modeling. Because the surrogate model that we are building is
for hybrid waveforms, we work with waveforms that are in the PN BMS frame.
However, while this ensures that the waveforms are in the same frame during the inspiral phase,\footnote{Formally, this only ensures that the waveforms are in the same BMS frame at $t\rightarrow-\infty$.} this is not necessarily true for the ringdown phase~\cite{MaganaZertuche:2021syq}. This is because, for example, the remnant black holes can have different kick velocities or supertranslation fields because of complicated effects that arise during the merger phase. Consequently, there are unresolved frame artifacts during the ringdown phase that can impact the modeling of the strain waveforms.

These additional challenges of modeling the strain in the ringdown using the current PN BMS framework are most pronounced for the $(3,1)$ and $(4,2)$ modes where the strain is oscillatory and does not decay to zero due to the impact of memory. Because these modes do not decay to zero as $t\rightarrow\infty$, we find that the decomposition of the strain into co-orbital frame data does not work well as the phase of these two modes is ill-defined. We highlight this phase issue in Fig.~\ref{fig:rd_plot}, which shows the amplitude and the phase of the $(3,1)$ mode for both a CCE waveform and an extrapolated waveform. One potential remedy to this problem is to instead work in a co-BMS frame, in which there is little-to-no time evolution of the BMS charges, i.e., a non-inertial frame similar to the co-rotating frame that simplifies the waveform data. However, such a project is non-trivial and we therefore postpone it for future work. 

\bibliography{References}

\end{document}